\documentstyle[12pt]{article}
\begin{document}
\setlength{\baselineskip}{0.27in}

\newcommand{\beq}{\begin{equation}}
\newcommand{\eeq}{\end{equation}}
\newcommand{\beqa}{\begin{eqnarray}}
\newcommand{\eeqa}{\end{eqnarray}}
\newcommand{\lsim}{\begin{array}{c}\,\sim\vspace{-21pt}\\<
\end{array}}
\newcommand{\gsim}{\begin{array}{c}\sim\vspace{-21pt}\\>
\end{array}}

\begin{titlepage}
{\hbox to\hsize{May 1994\hfill } }
{\hbox to\hsize{UM-TH-94-05\hfill hep-ph/9406217}}

\begin{center}
\vglue .06in
{\Large \bf Eikonal Wave Functions and Model Independent Exclusive B Decays}
\\[.5in]
\begin{tabular}{cc}
\begin{tabular}{c}
{\bf R. Akhoury\footnotemark[1]}\\[.05in]
{\it Physics Department}\\
{\it Brookhaven National Labs}\\
{\it Long Island, NY 11973}\\[.15in]
\end{tabular}
&
\begin{tabular}{c}
{\bf I.Z. Rothstein}\\[.05in]
{\it The Randall Laboratory of Physics}\\
{\it University of Michigan}\\
{\it Ann Arbor, MI 48109 }\\[.15in]
\end{tabular}
\end{tabular}
\footnotetext[1]{ On leave of absence from the Univ. of Michigan, Ann Arbor.}
\vskip 0.25cm
{\bf Abstract}\\[-0.05in]
\begin{quote}
In this letter we calculate the proper normalization for the B meson
eikonal valence wave function used in exclusive B decays. The normalization
appropriate for a hard scattering process renormalized at the scale
$\mu$ is determined by calculating
the short distance contribution to the purely leptonic decay.
\end{quote}
\end{center}
\end{titlepage}
\newpage
\section{Introduction}

Recently, it has been proposed that exclusive B meson decay amplitudes may be
calculated in a consistent fashion. In  \cite{asy} it was shown that
the form factor for the the B meson decay into a pion
may be written in the form
\beq
\label{eq:amp}
M(p,p^\prime)
=\int {d^4k\over{(2 \pi)^4}}
 \int_0^1 d\xi \phi_B(k) H(k,\xi,p,p^{\prime})\phi_{\pi}(\xi)~,
\eeq
where $\phi_B(k)$ and $\phi_\pi(\xi)$ are the $B$ and $\pi$ meson
valence wavefunction respectively, and $H$ is the hard scattering amplitude.
To make the calculation of the hard scattering amplitude consistent
it is necessary to factorize the infrared divergences coming from
soft and collinear gluon.  Once factorization has been proven the infrared
divergences
may be absorbed into the the wavefunction, which contains all the
long distance physics.

In \cite{asy} the divergences from soft gluon exchange were shown to
sum into an eikonal phase which may be absorbed into the B meson
wavefunction.
Physically, this summation is a consequence of the the fact that in the
low energy effective theory the heavy quark acts as a static color source.
Furthermore, the collinear divergences were shown to be cutoff due
to higher order Sudakov corrections \cite{sudakov}.
Once it is known that the infrared divergences are factorizable (in the
case of the soft divergences) or cutoff (in the case of the mass
singularities),
the next step in the justification of the use of (\ref{eq:amp}),
is to show that the two particle wave function dominates the
decay amplitude.
In the case of B decays,
the contribution from states with larger number of partons will
be down by $O(\Lambda/m_b)$.

In \cite{asy}
 the dependence of the wave-function on the heavy quark mass was left
undetermined.
In this letter we calculate the scale dependence of the universal
eikonal B meson wavefunction. This study is motivated by the fact
that wave function must be appropriately normalized. In general, the
normalization will depend upon the hard scattering amplitude which is
of course process dependent. To lowest order in $\alpha_s$
and $\Lambda / m_b$ the normalization is given
by \footnote{This normalization corresponds to $f_\pi=93~MeV$.}
\beq
\label{eq:lowestordernorm}
\int dk^{-} \hat{\phi}_B (k^-)=\frac{1}{2 \sqrt{3}}f_B,
\eeq
and we have defined
\beq
\hat{\phi}_B(k^-)=\int dk^+d^2k_{\bot} \phi_B(k).
\eeq
This normalization, defined through the purely leptonic decay, includes
contribution from short (compared to the QCD scale) wavelength
gluons which may also contribute to the hard scattering
amplitude. Thus, to avoid double counting the contribution from these
hard gluons, we would like to determine a scale dependent normalization for
the wave function. In other words, it is necessary  to extract
the short distance contributions to the normalization of the wave function
from the purely leptonic decay.

\section{The Eikonal Wave Function}

Figure 1 illustrates the generic structure for a B meson decay.
The B meson is composed of $n$ incoming partons, one of which, the b quark,
decays into a light quark. $h$ labels the hard scattering amplitude, and the
final blob represents the hadronization of the final state partons into
the final state of choice. For our purposes we will assume that the two
 parton state dominates the decay process. In general, this
 must be shown
to be due to some dynamics.
As discussed in the introduction, to  make the calculation of the
decay amplitude consistent it is necessary to show that the divergences due
to soft gluon exchange can be factorized and placed into the wave function.
The eikonal wave function is defined in a manner so as to absorb all
such soft divergences, as we will now show.

Consider the absorbtion/emission of a soft gluon from the heavy quark line.
Heavy quark effective field theory \cite{hqet}
 (HQET) tells us that the heavy quark will
be off shell by {\it at most} $O(\Lambda_{QCD})$. Thus, for gluon with
momenta satisfying the constraint $q^2<\Lambda_{QCD}~m_b$, we may use the
Feynman rules expressed  in figure 2. As in HQET, the heavy
quark propagator satisfies the constraint
\beq
\label{eq:hqetwavefunction}
{1\over{2}}(1+v\!\!\!\slash)h_v=h_v.
\eeq

Using these Feynman rules we may factor out all soft gluon exchanges
from  the hard scattering amplitude \cite{asy}. This can be seen by going to
the
rest frame of the heavy quark and resumming all the soft gluons
into a path ordered non-abelian eikonal phase defined by
\beq
\label{eq:eikphase}
U(A^0)=P~ exp(-ig\int^0_{- \infty}~ A^0(\lambda~n_0)d\lambda).
\eeq
In this definition $n_\mu$ is the unit four-vector in the
direction of the heavy quark four velocity.

By inserting $UU^{-1}$ into the matrix element relevant for the decay
under study, we may systematically remove the divergences due to soft
gluon exchanges to all orders by absorbing the eikonal phase into
the wave function. In this way we may write the gauge invariant
two particle wave function
as
\beq
\label{eq:wavefunction}
\psi_B(x)=<0 \mid T(\bar{b}^{free}(0)U(A^0)q(x))\mid
B(p)>_{\scriptstyle{L}(q)},
\eeq
where the interaction lagrangian ${\scriptstyle{L}(q)}$, is independent of
the $b$ quark field.
In the above a trace over color indices is assumed. Furthermore, the gauge
invariance
of the wave function is made manifest via the introduction of an additional
path ordered Schwinger phase. The contribution to the decay process from this
additional phase factor is higher twist and will thus be neglected.
This definition  incorporates the decoupling
of the heavy quark, in that there are no b quark loops contributing to
the amplitude.
By using this wave function the hard scattering amplitude becomes
manifestly free of soft divergences as a consequence of the fact
that for every diagram with a gluon exchange we must subtract the
same diagram only using the heavy quark Feynman rules defined in
figure 2.

\section{Normalization of the Eikonal Wave Function}

To make contact between physical quantities and the above mentioned
wave function, it is necessary to normalize the wave function to $f_B$,
up to some constant through the purely leptonic decay. The lowest order diagram
is shown in figure 3, giving the normalization
in eq. (2). However, to make this normalization process independent
we should calculate the next order correction so that we
may remove the contribution from gluons with momenta large
compared to the QCD scale. This entails calculating the corrections
to the difference between the full and eikonalized currents.
Here we present the leading order in $\Lambda_{QCD}/m_b$
calculations
for both the full
and eikonalized renormalized currents, as shown in figures 3a, 3b.
In the results presented below we always project onto the pseudoscalar
color singlet state.

The hadronic part of the amplitude is given by
\beq
\langle 0 \mid J_\mu^h \mid B(p)\rangle=\sqrt{2} i f_B p_\mu,
\eeq
Where we have defined
\beq
J_\mu^h=-i\bar{b}\gamma_\mu(1-\gamma_5)u.
\eeq
This amplitude may also be written as
\beq
\int {d^4k\over{(2\pi)^4}} \phi_B(k,\mu)H_\mu (p,k,\mu)
.\eeq
The hard scattering amplitude $H_\mu(p,k,\mu)$
is projected onto the pseudoscalar color singlet
channel. To lowest order it is given by (see figure 1)
\beq H_0(p,k)={N_c \over{\sqrt{6}}}4 i m_b ,
\eeq
in the rest frame of the meson. In this equation, as well as in the rest
of the paper, we do not distinguish between the mesonic and heavy
quark mass.

In this frame the one loop renormalized full vertex is given by
\beqa
\label{eq:fullcur}
{1\over{\sqrt{6}}}Z_l^{1/2} Z_h^{1/2}
Tr(\gamma_5 (p\!\!\slash + m_b)\langle 0 \mid J_\mu \mid {b}
\bar{u}
\rangle)
=\quad \quad \quad \quad \quad \quad\quad\quad\quad\quad\quad
\quad\quad\quad & &\nonumber \\
 {N_c\over{\sqrt{6}}}
{i \alpha_s m_b\over{3 \pi}} \left[-2ln{- k^2\over {m_b^2}}-8ln({2p\cdot k
\over {m_b^2}})+ 4  \right] . &&
\eeqa
The wave function renormalizations are given by
\beqa
\label{eq:Zs}
Z^{1/2}_l&=&1-{\alpha_s\over{6 \pi}}
\left[{1\over{\epsilon}}-\gamma-ln{-k^2\over{4\pi\mu^2}}-1
\right] \nonumber
\\
Z^{1/2}_h&=&1-{\alpha_s\over{6 \pi}}
\left[{1\over{\epsilon}}-\gamma-ln{m_b^2\over{4\pi\mu^2}}
+4ln{2p \cdot k \over{m_b^2}}+4 \right].
\eeqa

The one loop renormalized eikonal vertex is
\beqa
\label{eq:eikcur}
{1\over{\sqrt{6}}}Z_l^{1/2} Z_Q^{1/2}
Tr(\gamma_5 (p\!\!\slash + m_b) \langle 0 \mid J_\mu \mid {b}
\bar{u}
\rangle)
=\quad \quad\quad\
\quad\quad\quad
\quad\quad\quad
\quad\quad\quad\quad
\nonumber & & \\ {N_c\over{\sqrt{6}}} {i \alpha_s m_b\over{3 \pi}}
\left[{6\over{\epsilon}}-10\gamma
-2ln{- k^2\over {4\pi \mu^2}}-4ln({(2n\cdot k)^2
\over {4\pi \mu^2}})+ 14  \right] .& &
\eeqa
The eikonal wave function renormalization  is
\beqa
\label{eq:eZs}
Z^{1/2}_Q&=&
1-{\alpha_s\over{3 \pi}}\left[{1\over{\epsilon}}-2\gamma+1-ln{(2n \cdot k)^2
\over{4\pi\mu^2}}
\right].
\eeqa
Notice that neither (\ref{eq:fullcur}) nor (\ref{eq:eikcur})
 contain any double logs which are usually present due to
the
region of integration where the gluon is both soft and
collinear to the light quark.
The absence of these logs
 is a consequence of the fact that we are projecting onto the color singlet
state, which doesn't couple to long wavelength gluons.
Also, as seen from
(\ref{eq:eikcur}),
the eikonal vertex necessitates an infinite counter-term because the eikonal
current is no longer partially conserved.

The hard scattering amplitude $H_{\mu}(p,k,\mu)$
in the B meson rest frame is given by the difference between
the full and eikonal vertices
\beq
\label{eq:hsa}
H_{0}(p,k,\mu)=
i{N_c\over{\sqrt{6}}}4m_b\left[{\alpha_s\over{6\pi}}\left(
3ln{m_b^2\over{4\pi \mu^2}}-5+5\gamma\right)
\right].
\eeq
The collinear divergences cancel to all orders in the hard scattering amplitude
even though
they may come from hard gluons. This is simply because the light quark is
treated on the same
footing in both the full as well as eikonal theories. Furthermore, the two
theories
must match at $\mu=m_b$.

Combining the lowest order result with (\ref{eq:hsa}) leaves
\beq
i\sqrt{2}m_bf_B=
\int {d^4k\over{(2\pi)^4}}\phi_B(k,\mu){N_c\over{\sqrt{6}}}(4im_b)
\left[
1+{\alpha_s\over{6 \pi}}\left( 3ln {m_b^2\over{4\pi \mu^2}}-5+5\gamma \right)
\right].
\eeq
Thus, the leading order correction to the wave function is given by
\beq
f_B={1\over{2\sqrt{3}}}\psi_B(0,\mu)\left[ 1+{\alpha_s\over{6\pi}}
\left(3ln {m_b^2\over{4\pi \mu^2}}-5+5\gamma\right)\right],
\eeq
and explicitly shows the separation of the hard gluon contribution to
the decay process.

We may resum the leading logs using the renormalization group leaving
\beq
\label{eq:result}
 f_B ={\psi_B(0,\mu)
\over{2 \sqrt{3}}}\left[ \frac {\alpha(m_b)}{\alpha(\mu)} \right] ^{\frac
{-2}{11-2n_f/3}}.
\eeq
In this expression the wave function at the origin is independent
of the heavy quark mass.
Equation (\ref{eq:result}) may be used to rederive the result originally found
by Voloshin and Shifman \cite{vs} for the ratio of the B to D meson decay
constants
\beq
\label{eq:vs}
\frac {\displaystyle f_B} {\displaystyle f_D}= \frac {\displaystyle
\sqrt{m_c}} {\displaystyle \sqrt{m_b}} \left[ \frac {\alpha_s(m_b)}
{\alpha_s(m_c)}\right]^{(\frac {-2}{11-2 n_f/3})}.
\eeq
Where $n_f$ is 4 and the dependence on the heavy quark masses comes from
normalizing the states in a mass independent fashion.

We would like to point out that the eikonal wave function is a
lattice observable
\cite{cb}. This  is an important point given the fact that final state
interactions make hadronic decays inaccessible to lattice simulations.
Thus, we believe that the method discussed in \cite{asy} and above
is presently
the only model independent method for calculating exclusive decay amplitudes.
This is especially true for the case of purely hadronic decays, because
the lattice methods are rendered impotent due to final state
interactions. Moreover it should be stressed that like light-cone
wave functions, the eikonal wave function will satisfy the usual
Bethe-Salpeter type integral equation.

In conclusion it should be mentioned that the formalism,
in the form discussed above, does not take into account any
non-peturbative effects other than those in the eikonal wave function.
These contribution of these effects to the
decay of B hadrons into light mesons are presently under study.

\vspace{0.2cm}
\centerline{\bf Acknowledgements}
The authors benefitted from conversation with A. Falk and G. Sterman.
RA would like to thank the Max Planck Institute for Physics, Munich, especially
Prof. W. Zimmerman for hospitality and support.
IZR gratefully acknowledges the
Aspen Center
for Physics where part of this work was done.

\vspace{0.2cm}
\newpage
\centerline{\bf Figure Caption}

\noindent Figure 1: Spacetime picture of the semi-leptonic decay of a
B meson into some final hadronic state H.

\noindent Figure 2: Feynman rules for eikonal vertices.

\noindent Figure 3a: The full annihilation vertex.

\noindent Figure 3b: The eikonal annihilation vertex.

\end{document}